\documentclass[aps,twocolumn,prc,amssymb,amsmath,floatfix,superscriptaddress,nofootinbib]{revtex4}
\usepackage[dvipsnames,usenames]{xcolor}
\usepackage{graphicx}
\usepackage{mathptmx}
\usepackage{bm}

\newcommand{\be}{\begin{equation}}
\newcommand{\ee}{\end{equation}}
\newcommand{\bea}{\begin{eqnarray}}
\newcommand{\eea}{\end{eqnarray}}
\newcommand{\ba}{\begin{align}}
\newcommand{\ea}{\end{align}}

\newcommand{\rme}{{\rm e}}
\newcommand{\rmi}{{\rm i}}
\newcommand{\rn}{{\rm n}}
\newcommand{\rp}{{\rm p}}
\newcommand{\rmo}{{\rm o}}

\newcommand{\para}{\parallel}
\newcommand{\bfsf}[1]{\textsf{\textbf{#1}}}

\usepackage{setspace}

\begin{document}

\title{Proton superconductivity in pasta phases in neutron star crusts}

\author{Zhao-Wen Zhang}
\author{C. J. Pethick}
\affiliation{The Niels Bohr International Academy, The Niels Bohr Institute, University of Copenhagen, Blegdamsvej 17, DK-2100 Copenhagen \O, Denmark}
\affiliation{NORDITA, KTH Royal Institute of Technology and Stockholm University, Roslagstullsbacken 23, SE-106 91 Stockholm, Sweden}
\begin{abstract}

 In the so-called pasta phases predicted to occur in neutron star crusts, protons are able to move easily over large distances because the nuclear matter regions are extended in space.  Consequently, electrical currents can be carried by protons, an effect not possible in conventional crystalline matter with isolated nuclei.  With emphasis on the so-called lasagna phase, which has sheet-like nuclei, we describe the magnetic properties of the pasta phases allowing for proton superconductivity.  We predict that these  phases will be Type II superconductors and we calculate the energy per unit length of a flux line, which is shown to be strongly anisotropic.  If, as seems likely, the pasta structure is imperfect, flux lines will be pinned and matter will behave as a good electrical conductor and flux decay times will be long.  We describe some possible astrophysical manifestations of our results.

 \end{abstract}

\maketitle

\section{Introduction}

When the density of matter approaches that in atomic nuclei, the competition between Coulomb and surface energies can lead to the formation of phases with extended nuclei having rod-like or sheet-like form, the so-called pasta phases.  Such phases were first proposed in the context of matter with a relatively high proton concentration \cite{pasta,CJPRavenhall}, such as occurs in stellar collapse, and later studied for matter in neutron stars, which has a lower proton concentration  \cite{Lorenz,OyamatsuNS}.   Whether or not the pasta phases are more stable than phases with isolated nuclei depends on the nucleon--nucleon interaction and, e.g., with the SLy4 Skyrme interaction the pasta phases were found not to be the ground state \cite{DouchinHaensel}.  Parametric studies with a family of relativistic mean-field theory models indicate that the stability of the pasta phases appears to be correlated with the derivative of the symmetry energy with respect to density \cite{BaoShen}, and more detailed study of stability with improved nucleon--nucleon interactions is needed.  Many microscopic investigations of the properties of pasta phases have been carried out, but few of them consider matter with the low proton concentrations encountered in neutron stars.

The focus of this paper is the electrical and magnetic properties of the pasta phases, especially that with sheet-like nuclei (lasagna).  Interest in the electrical properties of these phases was sparked by the work of Ref. \cite{Pons}, where it was argued that the dearth of isolated X-ray sources with long periods could be explained if the pasta phases were poor electrical conductors as a result of scattering of electrons by the disorder of the nuclear matter.  Some support for a low electrical conductivity in the pasta phases was provided by quantum molecular dynamics (QMD) simulations \cite{HorowitzBerryBriggs}, which employed a formalism based on the idea of an effective impurity concentration. However, direct calculation of transport properties from QMD calculations of the static structure factor for the protons led to the conclusion that the electrical conductivity was not significantly different from that in phases with isolated nuclei \cite{NandiSchramm}.  A similar conclusion about the neutrino opacity had previously been arrived at in Ref. \cite{Alcain}.

A number of dynamical properties of the pasta phases have been investigated.  Long-wavelength, low-frequency collective modes, when matter is locally electrically neutral and there are no electrical currents  have been studied in Refs.\ \cite{ MartinUrban,KobyakovCJPJETP, DurelUrban, KobyakovCJPinprep}.   In Ref. \cite{Kobyakov2018}, the effect of entrainment of neutrons and protons was  considered and the possibility of flux lines in the pasta phases was mentioned.  In this paper we argue that the pasta phases will behave as Type II proton superconductors, in which magnetic flux threads the matter in the form of quantized flux lines, and we calculate the properties of these flux lines.  Consequently,  the pasta phases are likely to behave as good electrical conductors.
We shall argue that it is a good approximation to use a hydrodynamic description of the superfluid protons because the distance to which a magnetic field can penetrate is generally large compared with both the spacing between lasagna sheets and superfluid coherence lengths.  The properties of flux lines are shown to be strongly dependent on the angle between the flux line and the symmetry axis of the pasta.

This article is organized as follows.  Section II presents a general description of the lasagna phase, which is the phase we focus on because it is expected to be the most prevalent of the pasta phases.  Expressions for long-wavelength contributions to the proton and neutron superfluid currents are considered in Sec. III.  In Sec. IV, the properties of flux lines in the lasagna phase are considered, first  the case of a line perpendicular to the lasagna sheets and then for general directions.  Consequences of proton superconductivity are considered in Sec. V, and  concluding remarks are made in Sec. VI.

\section{General considerations}

We begin by giving estimates of some characteristic length scales.  After formation, neutron stars cool very quickly and to a first approximation they may taken to be at zero temperature.
The thickness of a sheet of lasagna, $\ell$, is given by \cite{pasta, CJPRavenhall}
\be
\ell\equiv 2r_N=\left( \frac{6}{\pi}\frac{\sigma u}{n_{\rm{pi}}^2e^2(1-u)^2}  \right)^{1/3}\,\,\,\,\,\,\,\,\,
\ee
\be
\approx    27  \left(  \frac{\sigma}{0.1\,{\rm MeV/fm}^2} \left(\frac{n_{\rm pi}}{n_s/20}\right)^{-2} \frac{u}{(1-u)^2}\right)^{1/3}\;\;\;{\rm fm},
\label{thickness}
\ee
where $\sigma$ is the surface tension of the interface between nuclear matter and the neutron liquid, $u$ is the fraction of space filled with nuclear matter, $e$ is the elementary charge,  $n_{\rm pi}$ is the proton density in the nuclear matter phase and $n_s=0.16$ fm$^{-3}$ is the saturation density of symmetric nuclear matter.  The average proton density is $n_{\rp}=n_{\rm pi}u$.   The lattice spacing of the periodic structure, $d$, is given by
\be
d=\frac{\ell}{u},
\ee
which for $u=0.5$, $\sigma=0.1$ MeV fm$^{-2}$ and $n_{\rm pi}=0.05 n_s$ \cite{OyamatsuNS} is approximately 44 fm.

The proton superconducting coherence length in bulk matter corresponding to that in the lasagna sheets is given by
\bea
\xi_{\rp}=\frac{\hbar v_\rp}{\pi \Delta_\rp}
\approx 8.3\left(\frac{1\, {\rm MeV}}{\Delta_\rp}   \right) \left( \frac{n_{\rm pi}}{n_s/20}  \right)^{1/3} {\rm fm},
\eea
where $v_\rp =p_\rp/m$ is the proton Fermi velocity, with  $p_\rp$ being the proton Fermi momentum, $\Delta_\rp$ the proton superconducting gap and $m$ the proton mass.\footnote{The small difference between the neutron and proton rest masses is unimportant for the considerations of the present article.}
In the lasagna phase, this length gives the range of correlations in directions lying in the plane of a lasagna sheet.

The London length, which determines the distance to which a uniform magnetic field outside the matter can penetrate into it, is given by
\bea
\lambda_{\rm pi}=\left(\frac{4 \pi n_{\rm pi}e^2}{m_\rp c^2}\right)^{-1/2}
\approx  81  \left( \frac{n_{\rm pi}}{n_s/20}  \right)^{-1/2} {\rm fm}.
\label{London}
\eea
Since the London length is much larger than the coherence length, bulk matter in the interior of the pasta phases will be a  Type II superconductor and one may anticipate that the pasta phases will also be.
We turn now to a long-wavelength description of nucleon currents in the pasta phases.

\subsection{Currents in the pasta phases}

 To describe  disturbances on length scales large compared with both the spacing of the lasagna sheets or spaghetti strands  $\sim r_N$ and the superfluid coherence length, which are of comparable magnitude, we use a hydrodynamic approach.   Superfluid currents involve gradients of the phase of the pair condensate amplitude, and to avoid having to consider in detail the local superfluid currents on length scales of order $r_N$ we shall work with coarse-grained averages of  the  phases of the condensate pair wave functions  of the protons, $2\phi_\rp$, and the neutrons,  $2\phi_\rn$.  Currents of nucleons also result from displacements of the structure of the pasta, which we denote by ${\bm u}(\bm r)$, and the corresponding velocity is $\dot{\bm u}$: these currents correspond to a normal component in the parlance of two-fluid models. The
total current densities of neutrons and protons are the sums of superfluid and normal contributions and, to lowest order in  $   {\bm\nabla} \phi_\rn$, ${\bm\nabla} \phi_\rp$ and $\dot{\bm u}$ may be written phenomenologically in the form \cite{MartinUrban,KobyakovCJPJETP, DurelUrban}
\be
{\bm j}_\rp=\frac{\bfsf{n}_{\rm pp}^{\rm s}}{m}\cdot \left({\bm\nabla} \phi_\rp-\frac{e}{c} {\bm A}\right) +    \frac{\bfsf{n}_{\rm pn}^{\rm s}}{m}\cdot {\bm\nabla} \phi_\rn +    {\bfsf n}_\rp^\rn\cdot \dot{\bm u}
\label{currentp}
\ee
and
\be
{\bm j}_\rn=\frac{\bfsf{n}_{\rm nn}^{\rm s}}{m} \cdot {\bm\nabla}\phi_\rn+    \frac{\bfsf{n}_{\rm np}^{\rm s}}{m}\cdot \left({\bm\nabla} \phi_\rp-\frac{e}{c} {\bm A}\right)  +    {\bfsf n}_\rn^{\rm n}\cdot \dot{\bm u},
\label{currentn}
\ee
where the superscripts s and n correspond to superfluid and normal components.
Because the proton carries charge, the phase of the proton condensate must occur in the combination ${\bm\nabla} \phi_\rp -e {\bm A}/c$ to satisfy gauge invariance.  Here, as in most of the article, we shall work in units such that $\hbar=1$.  The terms with $\bfsf{n}_{\rm pn}^{\rm s}$ take into account the fact that, due to neutron--proton interactions, the current of one species of nucleon is affected by the phase of the condensate wave function of the other species, an effect referred to as `backflow' or `entrainment'.
As pointed out in Ref.\@ \cite{MartinUrban}, the superfluid density tensor $\bfsf{n}_{\alpha\beta}^{\rm s}$, where $\alpha$ and $\beta$ are labels for the nucleon species, is anisotropic.  If the lasagna sheets are of uniform thickness, the superfluid density is diagonal in a coordinate system with one basis vector perpendicular to the sheets and two basis vectors in a pair of orthogonal directions lying in the plane of the sheets, and we shall use the superscript $\parallel$ to denote the former components and $\perp$ to denote the latter ones. When the sheets have no modulations in the perpendicular directions, $ n_\alpha^{\rn\perp}=0$.   Since in this paper we shall consider only the linearized theory, the superfluid and normal density tensors may be taken to be independent of the velocities.
Numerical simulations of lasagna for somewhat higher proton concentrations than those expected in neutron stars indicate that lasagna sheets are modulated in the $x$- and $y$-directions and that the modulations have hexagonal symmetry;  also for this case the density  tensors are isotropic for directions in the $xy$-plane but   $ n_\alpha^{\rn\perp}$ is nonzero.

From the requirement of Galilean invariance it follows that
\be
n_\alpha {\mathsf{\mathbf I}}={\bfsf n}_{\alpha}^\rn+\sum_\beta {\bfsf n}_{\alpha\beta}^{\rm s},
\label{Galinv}
\ee
where $\mathbf I$ is the unit tensor, $n_\rp$ is the proton density and $n_\rn$ is the neutron density.
This result was obtained for the case of pasta without spatial modulations in the plane of the sheets in Ref.  \cite{MartinUrban}, but it holds also when $ n_\alpha^{\rn \perp}\neq 0$.

In this article we shall consider situations where the pasta structure is stationary and the neutron superfluid phase is uniform, and therefore
\be
{\bm j}_\rp=\frac{\bfsf{n}_{\rm pp}^{\rm s}}{m}\cdot \left({\bm\nabla} \phi_\rp-\frac{e}{c} {\bm A}\right) .
\label{currentpsimple}
\ee
In terms of parallel and perpendicular components, the proton superfluid density tensor has the form
\be
{\bfsf n}_{\rm pp}^{\rm s}=n_{\rm pp}^{\rm s\perp} {\bfsf I}+(n_{\rm pp}^{\rm s\para}- n_{\rm pp}^{\rm s\perp})\hat{\bm\nu}\hat{\bm\nu},
\label{nstensor}
\ee
where $\hat{\bm\nu}$ is a unit vector perpendicular to the plane of the lasagna sheets.

In lasagna one expects the component of the proton superfluid density tensor $n_{\rm pp}^{\rm s\perp}$ to be of the order of the proton density, since protons can move relatively freely in the plane of the lasagna.  However, motion of protons perpendicular to the sheets is hindered by the layers of neutron matter between the nuclear matter regions.  To see this we estimate the rate at which protons at the Fermi surface in the neutron matter phase can tunnel through a layer of neutron matter.  We consider the most favorable situation for tunneling, in which the  incident neutron is moving normal to the interface. Inside the neutron matter, the  lowest energy state of a single proton is equal to its chemical potential, $\mu_{\rm po}$.
If one regards the neutron matter region as representing a rectangular barrier of height $\Delta \mu_\rp=\mu_{\rm po}-\mu_{\rm pi}$ above the proton chemical potential in the nuclear matter region, $\mu_{\rm pi}$, the magnitude of the tunneling amplitude is \cite{LLQM}
\be
|T|\simeq 4 \frac{(\Delta \mu_\rp \epsilon_{\rm pi})^{1/2}}{\Delta \mu_\rp +\varepsilon_{\rm pi}} \rme^{-\kappa\, \tau},
\label{tunnel}
\ee
for $\kappa \tau \gg1$, where $\kappa=(2 m\,\Delta \mu_\rp/\hbar^2)^{1/2}$ and $\tau=d-\ell$ is the thickness of a neutron matter layer.  The quantity $\varepsilon_{\rm pi} =(3\pi^2n_{\rm pi})^{2/3}/2m$ is the kinetic energy of a proton at the Fermi surface of the nuclear matter phase.
To obtain a rough estimate of $\Delta \mu_\rp$, we consider the difference between the proton chemical potentials for two phases in equilibrium at the highest pressure plotted in Ref.\@\cite[Fig. 2]{BBP}, where it is seen to be roughly $6$ MeV, and thus $\rme^{-\kappa \tau}\sim 3\times 10^{-7}$ for a neutron layer thickness equal to the thickness of a sheet, Eq. (\ref{thickness}). The prefactor in Eq. (\ref{tunnel}) is less than 2, so this cannot increase the order of magnitude of $|T|$, and we conclude that flow of protons between sheets is negligible when there are no ``bridges'' connecting adjacent sheets.
Even when there are bridges, it seems likely from simulations that their total cross sectional area per unit area of the sheet is much less than unity \cite{SchneiderBerryBriggsetalWaffles}.   Consequently, we expect $n_{\rm pp}^{\rm s\para}$ to be very much less than $n_{\rm pp}^{\rm s\perp}$ for the lasagna phase.  Since in the spaghetti phase protons can flow easily in the direction of the strands but only with difficulty between strands, $n_{\rm pp}^{\rm s\perp}$ will be much less than $n_{\rm pp}^{\rm s\para}$.

To begin with, we comment on properties of flux lines in an isotropic medium.  The flux line has a core, where superconductivity is suppressed, with a radius $\sim \xi_\rp$ and, outside this, superconducting currents extend out to a distance of order the London length.  As we shall show, this basic picture applies also to the pasta phases except that, because of the anisotropy of the screening length, the flux-line structure is also anisotropic. Flux lines in layered solid-state systems have been studied extensively \cite{Blatter}, but the pasta phases in neutron stars differ from layered superconductors in the laboratory  in that the superconducting regions have a thickness  much greater than the particle spacing.

\section{Flux lines in lasagna}

Uniform neutron star matter is expected to be a Type II superconductor, and consequently magnetic flux penetrating such matter will do so in the form of quantized flux lines \cite{BPP}. In lasagna, superconductivity is suppressed for currents perpendicular to the sheets because tunneling of protons between sheets is small.  We begin by considering the case of a single flux line perpendicular to the sheets in lasagna.

\subsection{Flux line perpendicular to the sheets}

To make explicit the physics of the situation, we shall analyse this problem in some detail.  It will turn out that the mathematics we shall use is well established in the context of terrestrial superconductors \cite{Tinkham}.

Within a lasagna sheet, a flux tube perpendicular to the lasagna sheets will tend to be compressed towards the center of the core by the circulating proton supercurrents.  In the regions between sheets, the flux will tend to  balloon outwards in order to reduce the magnetic field energy.    The local magnetic flux density, which we denote by ${\bm b}({\bm r}),$\footnote{In much of the literature on superconductivity, the symbol ${\bm h}({\bm r})$ is used for this quantity.  To avoid giving the impression that the local flux density is simply related to the magnetic intensity  $\bm H$, we prefer to use the symbol $\bm b$. The magnetic induction $\bm B$ is the spatial average of ${\bm b}({\bm r})$.} satisfies the Maxwell equation
\be
{\bm \nabla \times {\bm b} }=\frac{4\pi e{\bm j}_\rp}{c}.
\label{Maxwell}
\ee
The long-wavelength expression for the proton current, Eq. (\ref{currentpsimple}), is a good approximation provided the characteristic length scale is large compared with the coherence length, and therefore it can be used for most of the flux line but not for distances $\lesssim \xi_\rp$ from the center of the flux line.    We shall employ a coordinate system in which the $z$-axis is in the direction of the flux line, and the $x$- and $y$-directions are perpendicular to it.
Since the characteristic distance over which changes in the magnetic field can occur is of order the London length, Eq. (\ref{London}), which is much larger than the layer spacing, $b_z$ varies only weakly as a function of $z$ and, since ${\bm \nabla}\cdot {\bm b}=0$,   $b_x$ and $b_y$ are small compared with $b_z$.   Thus it is a good approximation to average Eq. (\ref{Maxwell}) over $z$. In the Appendix we give a specific example that illustrates this. As a simple model, we shall take the lasagna state to consist of sheets of uniform nuclear matter separated by neutron matter with a sharp boundary between the two sorts of matter.    For a flux line with a single quantum of flux directed in the positive $z$ direction, the phase of the proton pair condensate    $2 \phi_\rp$ changes by $2\pi$ when traversing a contour surrounding the center of the flux line but is zero otherwise.  Thus for a flux line centered at $x=y=0$ one may write $\bm\nabla \times \bm\nabla \phi_\rp=\pi \delta_2(\bm \rho)\hat{\bm z}$, where $\delta_2(\bm \rho)$ is the two-dimensional Dirac delta function at $\bm\rho= (x,y) =0$. In order not to complicate the notation, we shall denote the average magnetic induction by $\bm b$, and the curl of Eq. (\ref{Maxwell}) averaged over $z$ is
\be
   \frac{b_z}{{\lambda_\perp}^2} -\left(\frac{\partial^2} {\partial x^2} +\frac{\partial^2 }{\partial y^2}\right)b_z=\Phi_0  \delta_2(\bm \rho) .
   \label{bvortex}
\ee
Here
\be
\Phi_0=\frac{2\pi\hbar c}{2e}
\ee
is the flux quantum and
\be
\frac1{({\lambda_\perp})_0^2}= \frac{4 \pi (n_{\rm pp}^{\rm s \perp})_\rmi u  e^2}{mc^2},
\label{lambdaperpzero}
\ee
where the subscript $\rmi$ indicates that the quantity is to be evaluated for the matter in the interior of the nuclear matter sheets. The quantity $\lambda_\perp$ plays the role of an effective London length for screening of magnetic fields perpendicular to the plane of the lasagna sheets.  The subscript `0' indicates that the quantity is evaluated for the specific model of the lasagna we have employed;  more generally, when one uses the hydrodynamic equations for the coarse-grained average of the current,  Eqs. (\ref{currentpsimple}) - (\ref{nstensor}), the screening length is defined by
\be
\frac1{{\lambda_\perp}^2}= \frac{4 \pi n_{\rm pp}^{\rm s \perp}  e^2}{mc^2}.
\label{lambdaperp}
\ee
     The solution of Eq. (\ref{bvortex}) is
\be
{\bm b}(\rho)=\hat{\bm z}\frac{\Phi_0}{2\pi {\lambda_\perp}^2} K_0(\rho/\lambda_\perp),
\ee
where $\rho=(x^2+y^2)^{1/2}$ and $K_0$ is the modified Bessel function of the second kind.
For $\rho\ll \lambda_\perp$, the solution behaves as
\be
b(\rho)\simeq \frac{\Phi_0}{2\pi {\lambda_\perp}^2} \ln \frac{\lambda_\perp}{\rho},  \;\;\;\;\;\;          \xi_\rp \ll \rho \ll \lambda_\perp,
\ee
 to logarithmic accuracy.
Since the local approximation for the current density, Eq.\ (\ref{currentp}), fails within the core of the flux line at distances of order $\xi_\rp$ or less, we  cut the solution off at $\rho \approx \xi_\rp$.

The energy per unity length of the flux line may be calculated in the standard way \cite[Sec. 5.1.2]{Tinkham}, and is given by
\be
\epsilon \simeq \left( \frac{\Phi_0}{4\pi \lambda_\perp} \right)^2 \ln\frac{\lambda_\perp}{\xi_\rp}.
\ee
The energy per unit volume for flux lines with separations much greater than $\lambda_\perp$ and an areal density $n_v$ is therefore $E=n_v \epsilon$
and, since $\bm B =n_v \Phi_0{\hat {\bm z}}$, the corresponding value of the magnetic intensity  is given by
\be
{\bm H}=  4\pi\frac{\partial E}{\partial {\bm B}}\simeq {\hat{\bm z}} \frac{\Phi_0}{4\pi {\lambda_\perp}^2} \ln\frac{\lambda_\perp}{\xi_\rp}.
\label{H}
\ee
The component of $\bm H$ in the $x$- and $y$- directions vanishes by symmetry, as we shall show explicitly below.

\subsection{Flux line in a general direction}
We now calculate properties of a single flux line in a general direction, following the work of Refs. \cite{Grishin} and especially \cite{Balatskii}.  On multiplying Eq. (\ref{currentpsimple}) on the left by the matrix $({\bfsf n}_{\rm pp}^{\rm s})^{-1}$, and inserting  the curl of the result into the Maxwell equation (\ref{Maxwell}), one finds
\be
{\bm b}      +\frac{mc^2}{4\pi e^2}{\bm \nabla} \times \left(({\bfsf n}_{\rm pp}^{\rm s})^{-1}       {\bm \nabla} \times {\bm b}\right) = \hat{\bm l}\; \Phi_0\delta_2(\bm \rho),
\label{vortex}
\ee
where
$\hat{\bm l}$ is a unit vector in the direction of the flux line.
The inverse of the proton superfluid density tensor is
\be
({\bfsf n}_{\rm pp}^{\rm s})^{-1}=\frac {\bfsf I}{n_{\rm pp}^{\rm s\perp}}+\left(\frac1{n_{\rm pp}^{\rm s\para}}-\frac1{n_{\rm pp}^{\rm s\perp}}\right)\hat{\bm\nu}\hat{\bm\nu}.
\ee

 To solve Eq. (\ref{vortex}), it is convenient to take a Fourier transform in the plane perpendicular to $\hat{\bm l}$ and write
\be
{\bm b}_{\bm k}=\int d^2{\bm \rho}\,{\bm b}({\bm \rho})\exp(-\rmi {\bm k}\cdot {\bm \rho}).
\ee
 As pointed out by Takanaka, in anisotropic superconductors the local magnetic field has components perpendicular to the axis of the flux line \cite{Takanaka}.  We write $\bm b_{\bm k}$ in terms of its components along and perpendicular to $\hat{\bm l}$.  The perpendicular component is in the direction of ${\bm k}\times \hat{\bm l}$ since the component along $\bm k$ vanishes as a consequence of the Maxwell equation $\bm \nabla \cdot \bm b=0$. On solving the Fourier transform of Eq. (\ref{vortex}), one finds\footnote{Equation (\ref{bl}) agrees with Eq.\@(11) of Ref.\@ \cite{Balatskii} but there is a misprint in Eq.~(11$^\prime$) for ${\bm b}\cdot(\bm k \times \bm\nu)$, which should have the opposite sign.  This difference does not affect any of the other results in that paper.}
\be
\bm b\cdot \hat{\bm l}=  \Phi_0    \frac{1+k^2\lambda_\perp^2 \sin^2\theta    +k^2\lambda_\para^2 \cos^2\theta }{  (1+k^2\lambda_\perp^2)[1+({\bm k\cdot{\bm \nu}})^2\lambda_\perp^2 +({\bm k\times{\bm \nu}})^2)\lambda_\para^2]}
\label{bl}
\ee
and
\be
\bm b\cdot( \hat {\bm k}\times \hat{\bm l})=   \Phi_0    \frac{k^2(\lambda_\para^2 -\lambda_\perp^2) \cos\theta  {\bm \nu \cdot(\hat{\bm k}\times \hat{\bm l} )} }{  (1+k^2\lambda_\perp^2)[1+({\bm k\cdot{\bm \nu}})^2\lambda_\perp^2 +({\bm k\times{\bm \nu}})^2)\lambda_\para^2]},
\label{bperp}
\ee
where $\theta$ is the angle between the direction of the flux line and the normal to the lasagna sheets.  The screening length $\lambda_\para$ for magnetic fields lying in the plane of the sheets is defined analogously to Eq. (\ref{lambdaperp}):
\be
\frac1{{\lambda_\para}^2}= \frac{4 \pi n_{\rm pp}^{\rm s \para}  e^2}{mc^2}.
\label{lambdapara}
\ee
If $\bm \nu$ is taken to lie in the $xz$-plane, ${\bm \nu} \cdot(\hat{\bm k}\times \hat{\bm l})=\sin \theta k_y/k$.
The energy density associated with the flux line is the sum of the kinetic energy density, which may be written in the form
\be
E_{\rm kin}=\frac{m}2 {\bm j_\rp}\cdot    ({\bfsf n}_{\rm pp}^{\rm s})^{-1}   \cdot{\bm j_\rp}
\ee
when ${\bm \nabla} \phi_\rn=0$, and the magnetic energy density $b^2/(8\pi)$.  With the help of Eq.\ (\ref{vortex}) and integration by parts, one finds for the energy per unit length of the flux line
\be
\epsilon=   \frac{\Phi_0}{8\pi}\int \frac{ d^2{\bm k}}{(2\pi)^2} {\bm b}_{\bm k}\cdot \hat{\bm l}.
\ee
The integral here diverges logarithmically for large $\bm k$ if one uses Eq. (\ref{bl}) for ${\bm b}_{\bm k}$ because the structure of the flux core has not been treated in detail.  To take into account the effect of the core, it is therefore necessary to cut the integral off at wave numbers corresponding to the inverse of the core dimensions.  Detailed results for the energy per unit length of a flux will be given in the following subsection where, by working in coordinate space rather than Fourier space, we present a simple derivation of them.
\subsection{Simple derivation}

When the penetration depths are very much greater than the coherence lengths, the energy of a flux line may be calculated rather simply by making use of the fact that the kinetic energy of the currents dominates, and the energy of the magnetic field, which does not contain the logarithmic factor, is smaller.  At distances less than the characteristic screening lengths from the center of the flux line, the currents are essentially those of an uncharged vortex line, $\bm j_\rp={\bfsf n_{\rm pp}}\cdot {\bm \nabla}\phi_\rp$ since the vector potential is small there.   As earlier, we take the axis of the flux line to lie in the $z$ direction and $\hat {\bm \nu}$, the normal to the lasagna sheets, to lie in the $xz$-plane.  From the condition for translational invariance in the $z$ direction, the phase $\phi_\rp$ is a function only of $x$ and $y$.  In principle $\phi_\rp$ could have a contribution linear in $z$, which would correspond to a flow of protons along the flux line, but this situation is not of interest here.   Working with the superfluid momentum $\sim \bm\nabla \phi_\rp$, which has only two nonvanishing components is simpler than working with the proton current density, which has three.  In a steady state, the continuity equation for the proton current density is therefore
\be
\bm \nabla \cdot \bm j_\rp=(n_{\rm pp}^{\rm s})_{xx}\frac{\partial^2 \phi_\rp}{\partial x^2}+(n_{\rm pp}^{\rm s})_{yy}\frac{\partial^2 \phi_\rp}{\partial y^2}=0,
\ee
where
\be
(n_{\rm pp}^{\rm s})_{xx}=n_{\rm pp}^{\rm s\perp}\cos^2\theta+n_{\rm pp}^{\rm s\para}\sin^2\theta
\label{n_xx}
\ee
and
\be
(n_{\rm pp}^{\rm s})_{yy}=n_{\rm pp}^{\rm s\perp}.
\label{n_yy}
\ee
There is no term proportional to $\partial^2 \phi_\rp/\partial x \partial y$ because of our choice of axes.  One thus sees that in terms of the variables $X=x/\sqrt{(n_{\rm pp}^{\rm s})_{xx}}$ and $Y=y/\sqrt{(n_{\rm pp}^{\rm s})_{yy}}$ the phase $\phi_\rp$ satisfies Laplace's equation, and the solution corresponding to a singly quantized vortex is $\phi_\rp=\frac12\arctan(Y/X)$ and the streamlines are circles.

The kinetic energy density is
\be
E_{\rm kin}=\frac{1}{2m}\left( (n_{\rm pp}^{\rm s})_{xx}\left(\frac{\partial \phi_\rp}{\partial x}\right)^2+(n_{\rm pp}^{\rm s})_{yy}\left(\frac{\partial \phi_\rp}{\partial y}\right)^2  \right)
\ee
and therefore the energy per unit length of the flux line is
\be
\epsilon=   \frac{1}{8m} \left((n_{\rm pp}^{\rm s})_{xx}(n_{\rm pp}^{\rm s})_{yy}\right)^{1/2} \int \frac{dX dY}{X^2+Y^2},
\label{vortexenergy}
\ee
 where the integral is to be taken outside the core of the flux line out to distances at which the vector potential term in the expression for the current becomes significant. Introducing the radial coordinate $R=(X^2+Y^2)^{1/2}$, we may write the integral in Eq. (\ref{vortexenergy}) to logarithmic accuracy as
 \be
  \int \frac{dX dY}{X^2+Y^2}\simeq 2\pi  \ln\frac{R_>}{R_<},
 \ee
where $R_>$ is an upper cutoff radius and $R_<$ is a lower cutoff radius. The upper cutoff is determined by the smaller of the penetration depths $\lambda_x$ and $\lambda_y$ for magnetic fields varying in the $x$- and $y$-directions, measured in units of $[(n_{\rm pp}^{\rm s})_{xx}]^{1/2}$ and $[(n_{\rm pp}^{\rm s})_{yy}]^{1/2}$, respectively; since $\lambda_x \propto  [(n_{\rm pp}^{\rm s})_{yy}]^{-1/2}$ and $\lambda_y \propto  [(n_{\rm pp}^{\rm s})_{xx}]^{-1/2}$, the two ratios are identical.    The lower cutoff is determined by the structure of the core of the flux line.   The supercurrent is essentially confined to the lasagna sheets and in this plane, the flux-line core is isotropic and has a radius $\sim \xi_\rp$,  In the direction perpendicular to the lasagna sheets, the hydrodynamic approximation we have made will fail for distances comparable to the spacing between sheets.  Since the coherence length and the spacing of the sheets are of comparable magnitude, we adopt as a simple approximation a lower cutoff equal to $\xi_\rp$ in all directions. Thus   in both the $x$- and $y$-directions, the lower cutoff  is $\sim \xi_\rp$.   In the $XY$-plane the flow will cease to be circular at a radius that is the larger of  $\xi_\rp/[(n_{\rm pp}^{\rm s})_{yy}]^{1/2}$ and $\xi_\rp/[(n_{\rm pp}^{\rm s})_{xx}]^{1/2}$. Since for lasagna we expect $ n^{\rm s \perp}_{\rm pp}   \gg n^{\rm s \para}_{\rm pp}$, we see from Eqs.\@ (\ref{n_xx}) and (\ref{n_yy}) that the former is always smaller than the latter, which thus determines the cutoff. The energy per unit length of the flux line is then
\bea
&&\epsilon\simeq \frac{\pi}{4m} \left[(n_{\rm pp}^{\rm s})_{xx}(n_{\rm pp}^{\rm s})_{yy}\right]^{1/2} \ln\frac{\lambda_\perp}{\xi_\rp}\nonumber\\
 &=&\!\!\!  \frac{\pi n_{\rm pp}^{\rm s\perp} }{4m} (\cos^2\theta+g \sin^2\theta)^{1/2}\ln\frac{\lambda_\perp }{\xi_\rp  }.
\label{vortexenergy2}
\eea
The quantity
\be
g=\frac{n_{\rm pp}^{\rm s\para}}{n_{\rm pp}^{\rm s\perp}}
\ee
is a measure of the isotropy of the superfluid density tensor, $g=1$ corresponding to the isotropic case and $g=0$ to $n_{\rm pp}^{\rm s\para}=0$.  Implicit in the calculation to logarithmic accuracy is that the argument of the logarithm is large compared with unity.  Interestingly, the coefficient of the logarithm is independent of the elementary charge; this reflects the fact that the dominant contribution to the energy is due to the kinetic energy, which is essentially independent of $e$ for distances small compared with the screening length.

\begin{figure}
\includegraphics[width=3.5in]{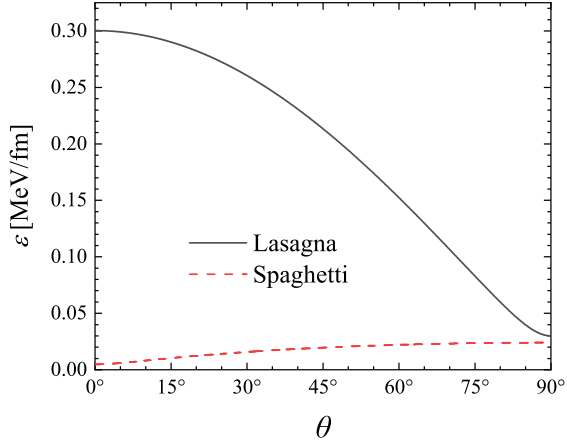}
\caption{Energy per unit length of a flux line in the pasta phases as a function of the angle between the flux line and the symmetry axis of the pasta.  The parameters employed are $n_{\rm pp}^{\rm s \perp}=0.025$  fm$^{-3}$, $n_{\rm pp}^{\rm s \para}=0.01n_{\rm pp}^{\rm s \perp}$ and $\lambda_\perp/\xi_\rp=10$ for lasagna, and $n_{\rm pp}^{\rm s \para}=0.020$  fm$^{-3}$, $n_{\rm pp}^{\rm s \perp}=0.01n_{\rm pp}^{\rm s \para}$ and $\lambda_\para/\xi_\rp=10$ for spaghetti.}
\label{energyperunitlength}
\end{figure}

The quantity $n^{\rm s \perp}_{\rm pp}$ is not well known even for uniform matter.  For that case, the normal component of the density is zero and the result (\ref{Galinv}) reduces to $  n^{\rm s \perp}_{\rm pp}=n_\rp-n^{\rm s \perp}_{\rm np}$.  Estimates of $n^{\rm s \perp}_{\rm np}$ have been made from various arguments based on Skyrme interactions and effective masses in neutron matter and symmetric nuclear matter.   These are uncertain but indicate that it is negative and  its magnitude is small compared with $n_\rp$ \cite{KPRS}.    If one neglects  in Eq.\ (\ref{Galinv}) both $n^{\rm s \perp}_{\rm np}$ and $n^ {\rn \perp}_\rp$, which is nonzero when band structure due to the spatial modulations of the pasta sheets is taken into account, one finds  $n^{\rm s \perp}_{\rm pp} \approx n_\rp$ as  a first estimate.  In Fig. \ref{energyperunitlength} we plot $\epsilon$ as a function of $\theta$ for lasagna for representative parameters.

Out basic formalism can also be applied to the spaghetti phase.  An important difference is that, in the spaghetti phase, superfluid flow of protons is possible in only one dimension (along the strands) in the absence of proton flow between strands.  In order to make current loops in this phase, proton flow between strands is necessary.   Whereas for lasagna   $ n^{\rm s \para}_{\rm pp}$ is very much less than $n^{\rm s \perp}_{\rm pp}$, the opposite is true for spaghetti.  Since $ n^{\rm s \perp}_{\rm pp}$ for lasagna and $ n^{\rm s \para}_{\rm pp}$ for lasagna  are of the order of the total proton density, one sees from Eqs. (\ref{vortexenergy2}) and (\ref{vortexenergyspaghetti}) that flux-line energies for spaghetti will be considerably smaller than those for lasagna.   On making arguments on the lower cutoff of the spatial integrals similar to those for the case of lasagna above, one finds for the energy per unit length of a flux line
\bea
&&\epsilon\simeq \frac{\pi}{4m} \left[(n_{\rm pp}^{\rm s})_{xx}(n_{\rm pp}^{\rm s})_{yy}\right]^{1/2} \ln\frac{\lambda(\theta)}{\xi_\rp}\nonumber\\
 &=&\!\!\!  \frac{\pi [n_{\rm pp}^{\rm s\perp}  (n_{\rm pp}^{\rm s\perp}\cos^2\theta+n_{\rm pp}^{\rm s\para} \sin^2\theta)]^{1/2}}{4m}\ln\frac{\lambda(\theta) }{\xi_\rp  },
\label{vortexenergyspaghetti}
\eea
where
\be
\lambda(\theta)^{-2}=\lambda_\perp^{-2}  \cos^2\theta+ \lambda_\para^{-2}\sin^2\theta.
\ee
The energy of a flux line per unit length is proportional to $n^{\rm s \perp}_{\rm pp}$   for $\theta=0$ and increases with increasing $\theta$ until for $\theta=\pi/2$ it is proportional to $(n^{\rm s \para}_{\rm pp}n^{\rm s \perp}_{\rm pp})^{1/2}$.  In Fig. \ref{energyperunitlength} we plot the energy per unit length of a flux line in spaghetti as a function of angle for $ n^{\rm s \para}_{\rm pp}=0.02 n_{\rm s}$ and $ n^{\rm s \perp}_{\rm pp}/n^{\rm s \para}_{\rm pp}=0.01$.

 The magnetic intensity for the lasagna phase is given by the first part of Eq. (\ref{H}) and Eq. \ref{vortexenergy2}, and to logarithmic accuracy has the form
 \bea
 &&{\bm H}\!\!\simeq \frac{\Phi_0}{4\pi {\lambda_\perp}^2} \ln\frac{\lambda_\perp}{\xi_\rp}    \label{Hgen}\\
 &\times&\!\!\!  \left( (\cos^2\theta +g \sin^2 \theta)^{1/2}{\hat{\bm z}} + \frac{(1-g)\sin \theta \cos \theta}{(\cos^2\theta +g \sin^2 \theta)^{1/2}} \hat{\bm x}      \right), \nonumber
 \eea
 for a flux line lying in the $xz$-plane.   For $n^{\rm s \para}_{\rm pp}=0$ ($g=0$), Eq. (\ref{Hgen}) reduces to
 \be
 {\bm H}\simeq     \frac{\Phi_0}{4\pi {\lambda_\perp}^2} \ln\frac{\lambda_\perp}{\xi_\rp}  \hat{\bm l} \,{\rm sgn}(  \hat{\bm \nu} \cdot \hat{\bm l} ),
 \label{Hgen0}
 \ee
 which lies in the direction of the normal to the lasagna sheets and, apart from the logarithm, is independent of the angle $\theta$ between the flux line and the normal to the lasagna sheets: the energy of a flux line is proportional to the length of the projection of the line onto the normal to the lasagna sheets.   For an isotropic superconductor, $H \simeq 4\pi \epsilon/\Phi_0$ but for an anisotropic one it can be very much larger.

 The magnetic intensity calculated above represents the lowest value for which a flux line can be present in the medium, for a given angle between the symmetry direction of the pasta and the magnetic induction, and therefore corresponds to $H_{\rm c1}$ in the language of condensed matter physics.  Numerically, for $\theta=0$ one finds
 \be
 H_{\rm c1}(\theta=0)\approx 1.27 \times 10^{14} \left(\frac{n^{\rm s \perp}_{\rm pp}}{n_{\rm s}/40}  \right) \ln\frac{\lambda_\perp}{\xi_\rp}\;\;\; {\rm Oe} .
 \ee
 The ratio ${\lambda_\perp}/{\xi_\rp}$ is of order 10, so $ H_{\rm c1}(\theta=0)\sim 3 \times 10^{14}$ Oe, somewhat lower than its value in the uniform interior  \cite{BPP} because of the lower proton density in the crust.

 With increasing $\bm B$, flux lines will arrange themselves into a periodic array.  Calculations in Ref. \cite{Balatskii} indicate that the array is made up of isosceles triangles, rather than the equilateral ones found for isotropic superconductors.  This picture will be made more complicated when one takes into account the  periodicity of the pasta structure, since there will be two competing length scales, one associated with the spacing of the lasagna sheet and the other with the typical spacing between flux lines.  As in other condensed matter problems, one can thus expect a series of transitions between different configurations \cite{Bak}.

 Superconductivity is suppressed completely at the upper critical field, $H_{\rm c2}(\theta)$, which is given in the anisotropic Ginzburg--Landau model by \cite{Morris, Takanaka2}
 \be
 H_{\rm c2}(\theta)=\frac{\Phi_0}{2\pi \xi_{\rp \perp} \xi(\theta)},
 \ee
 where
 \be
 \xi(\theta)^2=  \xi_{\rp\perp}^2 \cos^2\theta+ \xi_{\rp\para}^2 \sin^2\theta.
 \ee
 Here $\xi_{\rp\perp}$ and $\xi_{\rp \para}$ are the proton coherence lengths perpendicular to and parallel to the symmetry axis of the pasta.  If we take the two coherence lengths to be $\sim \xi_\rp$, the predicted value of $H_{\rm c2}$ is of order $3 \times 10^{16}$ Oe, similar to that in bulk matter in the liquid interior \cite{BPP}.
 At $H_{\rm c2}$, $\bm B$ and $\bm H$ have the same magnitude and direction.   We caution the reader that the anisotropic Ginzburg--Landau model does not take into account the periodic structure of the pasta phases,  which can have an important effect, since the lattice vectors of these phases is comparable in magnitude to the superfluid coherence length.

   \section{Consequences of proton superconductivity}

  In this section, we explore a number of ways in which proton superconductivity in the pasta phases would manifest itself in neutron stars.

  \subsection{Torques}  The fact that the energy of a flux line depends on its direction with respect to the lasagna layers implies that there are forces tending to make the flux line lie in the plane perpendicular to the normal to the lasagna sheets.  The alignment torque on unit length of the flux line is
 \be
  {\bm \Gamma}=    \frac{\partial \epsilon }{\partial \cos \theta}           \hat{\bm \nu}\times  \hat{\bm l},
 \ee
and the torque on the pasta structure is the opposite of this. These torques will influence the frequencies of collective modes of magnetized pasta phases.

The dependence of the flux-line energy on orientation has implications for the relative orientation of pasta phases with respect to the magnetic field in newly born neutron stars.  If the pasta phases are formed at temperatures lower than the transition temperature for proton superconductivity, the lasagna sheets will tend to form so that the magnetic induction lies in the plane of the sheets, since this configuration has lower (free) energy.
If the proton superconducting transition temperature lies below the temperature for pasta formation, there will be no correlation between the the direction of the magnetic field and the orientation of the pasta.

 \subsection{Magnetic field evolution}

 In a new-born neutron star, protons are normal before the temperature falls to the superconducting transition temperature and  magnetic fields are supported by electrical currents produced by electrons.  When the temperature falls below the superconducting transition temperature, the transition will occur at constant $\bm B$, since the time to move flux over distances typical of neutron star dimensions is very large \cite{BPP,BaymNordita}. Thus proton currents will be induced in the regions where protons are superconducting, thereby collecting the flux into quantized flux lines.  The rate at which the magnetic field in the superconducting regions can evolve depends on the forces on the flux lines, and how rapidly energy can be dissipated due to motion of a line.

Determining how the magnetic field evolves after the protons become superconducting requires that one take into account the fact that the pasta phases are unlikely to be perfectly ordered, in which case there will be forces on flux lines which will pin them in place, as in terrestrial Type II superconductors \cite{Blatter}.    Because the energy per unit length of a flux line depends strongly on its direction relative to the lasagna sheets, the pinning forces will be large.   To the extent that flux lines are pinned, flux will be locked in place and the flux density will remain constant in time.  Even if currents in other parts of the star decay, large-scale magnetic fields could be supported by the persistent proton currents circulating in flux lines.

Even in perfectly ordered pasta phases, the proton superfluid properties are not homogeneous and vary on the scale of the spacing between layers, $d$ and the modulations of the structure within layers; this effect was not taken into account in our treatment, where averages were performed over such length scales.   This will result in a modulation of the flux-line energy on these length scales but its magnitude is of order $(d/\lambda_\perp)^2$ compared with the energy per unit length of the line, and therefore small.
Flux lines are subject to forces due to
gradients of the matter properties, of the density of flux lines, and of the angle between the flux lines and the normal to the lasagna sheets..  The rate at which flux lines can move with respect to the matter depends on the rate of energy dissipation by a moving flux line, a topic we shall defer to future work.

\section{Concluding remarks}

In this article we have shown that superconductivity of protons in the pasta phases in neutron star crusts will change the magnetic behavior qualitatively compared to that for phases with isolated nuclei. This is a consequence of the fact that, in these phases, the neutron matter component is spatially extended, rather than in isolated nuclei, as in ordinary matter.   The matter is predicted to be a Type II superconductor, in which magnetic flux occurs in quantized flux lines.    If flux lines are pinned, the lasagna phase will behave as a good electrical conductor and proton supercurrents there can sustain magnetic fields in the star.

 \section*{Acknowledgments} We are grateful to Alexander Balatsky for helpful correspondence.  Author Z.-W. Z.  is grateful to Nordita for support while this work was carried out.  CJP is grateful to Dmitry Kobyakov for helpful comments in the early stages of this work.

\appendix

\section{Microscopic structure of magnetic fields in lasagna}

Here we consider the special case of a magnetic field parallel to the lasagna sheets, which we take to lie in the $xy$-plane.  In the neutron matter between sheets, the magnetic induction is uniform, and we shall take it to be in the $x$-direction.  We shall assume that there is no flow of protons between sheets.  Inside the nuclear matter, the  magnitude of the proton pair amplitude may be taken to be uniform because the coherence length is much greater than the thickness of the sheets.  In that region the local magnetic flux density, which we denote by ${\bm b}({\bm r})$, is a function only of $z$, the coordinate perpendicular to the plane of the lasagna sheets, and satisfies the London equation
\be
\frac{\partial^2 b_x}{\partial z^2}-\frac{b_x}{\lambda_{\perp \rmi}^2}=0,
\ee
where
\be
\frac{1}{\lambda_{\perp \rmi}^2}=\frac{4\pi (n_{\rm pp}^{\rm s\perp})_{\rmi}e^2}{mc^2}.
\ee
Here the subscript $\rmi$  on $(n_{\rm pp}^{\rm s\perp})_{\rmi}$ indicates that the superfluid density tensor is that for the nuclear matter phase. The solution is
\be
b_x(z)=b_{\rm o}\frac{\cosh(z/\lambda_{\perp\rmi})}{\cosh(d/2\lambda_{\perp\rmi})},
\ee
where $b_{\rm o}$ is the magnetic induction in the neutron matter regions.  The magnetic induction is the spatial average of $\bm b$, and is given by
\bea
B_x=b_\rmo \left[ 1-\frac{2r_N}{d} +2\frac{\lambda_{\perp\rmi}}{d}\tanh\left(r_N/\lambda_{\perp\rmi}\right)  \right]\\
\simeq b_\rmo\left( 1-\frac13 \frac{r_N^2}{{\bar\lambda}^2}  \right),
\eea
where the latter expression applies for $r_N\ll \bar\lambda$, and $\bar\lambda$ is given by
\be
{\bar\lambda}=\lambda_{\perp\rmi} \left(\frac{d}{\ell}\right)^{1/2},
\ee
which is the expression for $\lambda_\perp$, Eq. (\ref{lambdaperp}), when $n_{\rm pp}^{\rm s \perp}=u (n_{\rm pp}^{\rm s \perp})_\rmi$, its value for the model where the boundary between the nuclear matter and neutron matter is sharp. We therefore conclude that the variations in the magnetic field as a function of $z$ are of order $(d/\lambda_\perp)^2 \sim 10^{-2}$ times the mean field, and we expect this result to hold also for other directions of the average field and for arbitrary directions of the magnetic flux.

In the case considered here, there are no flux lines, since we have assumed there to be no flow of protons between lasagna sheets.  Matter behaves as a diamagnet because of the partial expulsion of flux from superconducting regions:  For small $\bm B$, $H$ is proportional to $B$ with a coefficient that is unity to within a correction of order  $(d/r_N)^2$.

When flow of protons between sheets is possible and for a magnetic field in the $x$-direction, flux will be bunched into flux lines with an extent $\sim \lambda_\perp$  in the $z$-direction and  $\sim \lambda_\para$ in the $y$-direction.

\end{document}